# Predictive Analysis for Social Processes I: Multi-Scale Hybrid System Modeling

Richard Colbaugh    Kristin Glass

*Abstract*—This two-part paper presents a new approach to predictive analysis for social processes. In Part I, we begin by identifying a class of social processes which are simultaneously important in applications and difficult to predict using existing methods. It is shown that these processes can be modeled within a multi-scale, stochastic hybrid system framework that is sociologically sensible, expressive, illuminating, and amenable to formal analysis. Among other advantages, the proposed modeling framework enables proper characterization of the interplay between the *intrinsic* aspects of a social process (e.g., the "appeal" of a political movement) and the *social dynamics* which are its realization; this characterization is key to successful social process prediction. The utility of the modeling methodology is illustrated through a case study involving the global SARS epidemic of 2002-2003. Part II of the paper then leverages this modeling framework to develop a rigorous, computationally tractable approach to social process predictive analysis.

## I. INTRODUCTION

ENORMOUS resources are devoted to the task of predicting the outcome of social processes, in domains such as economics, public policy, popular culture, and national security, but these predictions are often woefully inaccurate. Consider, for instance, the case of cultural markets. Perhaps the two most striking characteristics of these markets are their simultaneous *inequality*, in that hit songs, books, and movies are many times more popular than average, and *unpredictability*, so that well-informed experts routinely fail to identify these hits beforehand. Examination of other domains in which the events of interest are outcomes of social processes reveals a similar pattern – market crashes, regime collapses, fads, and "emergent" social movements involve significant segments of society but are rarely anticipated.

It is tempting to conclude that the problem is one of insufficient information. Clearly winners are qualitatively different from losers or they wouldn't be so dominant, the conventional wisdom goes, so in order to make good predictions we should collect more data and identify these crucial differences. Research in the social and behavioral sciences calls into question this conventional wisdom and, indeed, indicates that there may be fundamental limits to what can be predicted about social systems. Consider social processes in which individuals pay attention to what others do. Recent empirical studies offer evidence that the *intrinsic* characteristics of such processes, such as the quality of the various options in a social choice situation, often do not possess much predictive power. For example, the study reported in [1] finds that the (intrinsic) attributes ordinarily considered to be important when assessing the likelihood of movie box office success, such as the actors, genre, and critic reviews, are not statistically significantly related to box office revenue. Similar results hold for the adoption of innovations [e.g., 2,3], diffusion of social and political behaviors [e.g., 4,5], sales in various markets [e.g., 5-7], and the rise and fall of fads and fashions [e.g., 8]. Available experimental evidence, although scarcer, supports this conclusion [e.g., 9].

This research provides compelling evidence that, for many important social processes, it is not possible to obtain useful predictions using standard methods, which focus almost exclusively on the intrinsic characteristics of the process and its possible outcomes. We propose that accurate prediction, if achievable at all, requires careful consideration of the interplay between the intrinsics of the process and the underlying *social dynamics* which are its realization. This two-part paper presents a new approach to predictive analysis which exploits this idea. In Part I, we propose a multi-scale, stochastic hybrid system modeling framework for social processes that captures the interplay between a process' intrinsic features and its dynamics in a sociologically-grounded way. Then, in Part II, we formulate predictive analysis questions in terms of social dynamics reachability and present a rigorous, computationally tractable method for deciding reachability and, consequently, for answering prediction questions.

Taken together, this two-part paper makes three main contributions. First, we identify a class of social processes that are important in applications and difficult to predict using existing techniques, and we develop a multi-scale modeling framework for these processes. The proposed framework is sociologically-grounded, captures and illuminates important social structures, and provides a mathematical representation which supports formal analysis. Second, we derive a mathematically rigorous, computationally tractable approach to predictive analysis that consists of four capabilities: predictability assessment, identification of observables with predictive power, warning analysis, and prediction. Finally, the potential of the proposed approach is illustrated through case studies involving epidemics, online markets, social movements, and mobilization/protest behavior.

The research described in this paper was supported in part by the U.S. Department of Homeland Security and Sandia National Laboratories.

R. Colbaugh is with Sandia National Laboratories, Albuquerque, NM 87111 USA and New Mexico Institute of Mining and Technology, Socorro, NM 87801 USA (phone: 505-603-1248; e-mail: colbaugh@nmt.edu).

K. Glass is with New Mexico Institute of Mining and Technology, Socorro, NM 87801 USA (e-mail: kglass@icasa.nmt.edu).

## II. MULTI-SCALE MODELING FRAMEWORK

### A. Positive externality social processes

In many social situations, people are influenced by observations of (or expectations about) the behavior of others, for instance seeking to obtain the benefits of coordinated actions, infer otherwise inaccessible information, or manage complexity in decision-making. Processes in which observing a certain behavior increases an individual's probability of adopting that behavior are often referred to as *positive externality processes* (PEP), and we use that term here. PEP have been widely studied in the social and behavioral sciences and, more recently, by the informatics and systems theory communities.

One hallmark of PEP is their apparent unpredictability: phenomena from fads and fashions to financial market bubbles and crashes to political revolutions appear resistant to predictive analysis (although there is no shortage of *ex post* explanations for their occurrence!). It is not difficult to gain an intuitive understanding of the basis for this unpredictability. Individual preferences and opinions are mapped to collective outcomes through an intricate, dynamical process in which people react individually to an environment consisting largely of others who are reacting likewise. Because of this feedback dynamics, the collective outcome can be quite different from one implied by a simple aggregation of individual preferences; standard prediction methods, which typically are based (implicitly or explicitly) on such aggregation ideas, do not capture these dynamics and therefore are often unsuccessful.

### B. Multi-scale PEP models

In what follows, we focus on binary decision PEP, in which individuals choose between two options, A and B, and are more likely to choose option A if they observe others selecting A. This class of social dynamics is important in applications: people frequently are faced with a choice between two options and are motivated to behave as others do. Moreover, the case of finitely many options is essentially identical analytically [10].

To gain a quantitative understanding of the social process by which preferences and opinions of individuals become the collective outcome for a group, we model PEP in a manner which explicitly separates the individual, or "micro", dynamics from the collective dynamics. Specifically, we adopt a "multi-scale" perspective and develop a class of models for PEP which employs three modeling scales:

- a *micro-scale*, for modeling the behavior of individuals;
- a *meso-scale*, which represents the collective dynamics within "social contexts" via simple models for the interaction dynamics;
- a *macro-scale*, which characterizes the interaction between the social contexts.

Social contexts are localized social settings, determined by vocational organization, family structure, or physical neighborhood, in which interactions between individuals can be modeled as "fully mixed" – that is, all pairwise interactions between individuals within a social context are equally likely. This conceptualization is well-grounded in the social sciences (see [4,11,12] for social network-oriented discussions). Indeed, one advantage of identifying a scale at which agent interaction is (approximately) homogeneous is that this enables the leveraging of an extensive literature on collective dynamics. By adjusting the definition of social context it is possible to recover other social system representations, including compartment (context equals entire group) and agent-based (context equals individual) models. A schematic of the basic framework is given in Figure 1.

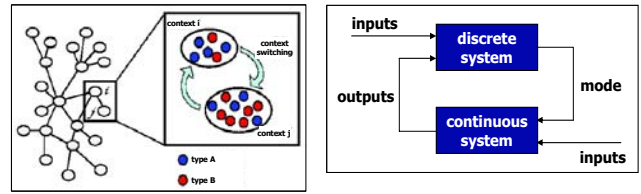

Fig. 1. Multi-scale model for social processes. The cartoon at left illustrates the basic model structure, in which individuals (blue and red nodes) interact *within* social contexts (ellipses encircling nodes) via fully mixed dynamics and *between* contexts according to the network topology characterizing context interdependency. The block diagram at right depicts a stochastic hybrid system encoding of the model, in which the continuous system captures intra-context dynamics and the discrete system models inter-context interactions.

We begin with a description of the micro scale. The micro-scale model quantifies the way individuals combine their own inherent preferences regarding the available options with their observations of the behaviors of others to arrive at their chosen courses of action. We seek a model which is consistent with empirical data and reflects current social science thinking concerning PEP. In particular, we wish to develop a micro-scale model that accommodates two of the main drivers for positive externality behavior:

- *utility-oriented externalities*, in which the utility or value of an option is a direct function of the number of others making that choice (see Example 2.1 below);
- *information externalities*, which arise from inferences made by an individual about decision-relevant information possessed by others (see Example 2.2 below).

Consider a binary choice setting, in which N agents choose from a set $O = \{0,1\}$ of options based in part on the choices made by others. Let $o_j \in \{0,1\}$ denote the selection of agent j and $o = [o_1 \ldots o_N]^T \in O^N$ represent the vector of choices made by the group. It is reasonable to suppose that agent i chooses between the options probabilistically according to some map $PO_i: A_i \times O^N \to [0,1]$, where $PO_i$ is the probability that agent i chooses option 1, $A_i$ measures i's in-herent preference for option 1, and $PO_i$ is nondecreasing in $A_i$. In positive externality situations $PO_i$ also should be "nondecreasing in o" in some sense, and we now make this notion precise. (For notational simplicity in what follows we often suppress the dependence of $PO_i$ on $A_i$.)

Because it is defined in such general terms, it may appear that the map $PO_i$ could be a very complicated function of the choices of the other agents. In fact, it can be shown that this map must be tractable.

**Proposition 1:** Given any $PO_i$ there exists a vector $w_i = [w_{i1} \ldots w_{iN}]^T \in \Re^N$, with $w_{ij} \geq 0$ and $\Sigma_j w_{ij} = b_i$, and a scalar function $r_i: [0, b_i] \to [0,1]$ such that $PO_i(o) = r_i(o^T w_i)$.

**Proof:** It is enough to prove that the $w_{ij}$ can be chosen so $o^T w_i: O^N \to [0, b_i]$ is injective, since then $r_i$ can be constructed to recover any $PO_i$. One such choice for $w_i$ is $w_i = [2^0 \ 2^1 \ \ldots \ 2^{N-1}]^T$, as then $o^T w_i$ provides a unique (binary number) representation for each o. ∎

We call $r_i$ the *agent decision function* and $s_i = o^T w_i$ agent i's *social signal*, and interpret the $w_{ij}$ as defining a weighted social network for the group of N agents. The requirement that the $w_{ij}$ be nonnegative implies that an increasing social signal $s_i$ corresponds to an increasing weighted fraction of i's neighbors choosing option 1. Observe that Proposition 1 quantifies the way social influence is transmitted to an agent by her neighbors and highlights the importance of this signal in the decision-making process. The result also allows a simple characterization of positive externality agent behavior: for such behavior, $r_i$ is nondecreasing in $s_i$.

We now show that this model structure easily accommodates both utility-oriented and information externalities.

**Example 2.1: Utility-oriented externalities**

Suppose each agent i has a utility function $u_i: O \times [0, b_i] \to \Re^+$ which depends explicitly on i's social signal $s_i$. The key quantity considered by agent i when selecting between options 0 and 1 is the utility difference between the options, $\Delta u_i(s_i) = u_i(1,s_i) - u_i(0,s_i)$. Note that in positive externality situations $\Delta u_i$ is increasing in $s_i$. Thus there exists a *threshold* social signal value $s^*$, possibly with $s^* < 0$ or $s^* > b_i$, such that a utility maximizing agent is more likely to choose option 0 if $s_i < s^*$ and option 1 if $s_i \geq s^*$. For instance, in models of technology adoption [e.g., 2], utility functions of the form $u_i(o,s_i) = a_i(o) - p(o) + e(o,s_i)$ are often proposed, where $a_i$, p, e reflect intrinsic utility, price, and social effects, respectively. Then, in the usual case in which $e(1,s_i)$ ($e(0,s_i)$) increases (decreases) with $s_i$, we have $d(\Delta u_i)/ds_i > 0$ and externalities are positive.

**Example 2.2: Information externalities**

Suppose the utility to agent i of each option is independent of the number of other agents choosing that option but there exists uncertainty regarding this utility. To be concrete, assume that agent i's utility depends on the "state of world" $w \in \{w_0, w_1\}$, so that $u_i = u_i(o,w)$, and there exists uncertainty regarding w. In this case, agent i may observe others' decisions in order to infer w and then choose the option which maximizes his utility for this world state. Consider, for example, the decision of whether to adopt an innovation of uncertain quality, and let the world state $w_1$ signify that innovation quality is such that adopting maximizes utility. In this situation it is reasonable for agent i to maximize *expected* utility and choose the option (adopt or not) $o^* = \arg\max_{o \in O} \Sigma_{w \in W} P(w \mid s_i) u_i(o,w)$. If agent i uses Bayesian inference to estimate $P(w_1 \mid s_i)$ then we have a positive externality decision process and there exists a threshold value $s^*$ for the social signal such that agent i is more likely to choose option 0 if $s_i < s^*$ and option 1 if $s_i \geq s^*$ [10].

In each of the preceding examples, agents exhibit a threshold behavior: agent i is more likely to choose option 1 if the weighted fraction of i's neighbors choosing this option is large enough. This sort of behavior also arises in many other social theoretic settings (e.g., coordination games). We remark that, by explicitly representing the individual agents' decision processes via decision functions $r_i$, the proposed micro-scale model supports empirical estimation of this function in some circumstances [10].

Consider next the meso-scale component of the proposed multi-scale modeling framework. Taken together, the meso-scale and macro-scale models for social processes quantify the way agent decision functions interact to produce collective behavior at the scale of large groups (e.g., a segment of society). The role of the meso-scale model is to quantify and illuminate the manner in which agent decision functions interact *within* social contexts, while the macro-scale model characterizes the interactions of agents *between* contexts.

We first introduce a simple formulation for fully mixed (meso-scale) collective dynamics and then employ this foundation to derive a few useful models for positive externality dynamics. Consider a population of individuals with agent decision functions $r_i$ and interaction structure defined by $w_{ij}$. Assume that these agents form a single social context, and that within this context agent interaction is global and anonymous, $w_{ij} \equiv w > 0$. Then each agent's decision function $r_i: [0,1] \to [0,1]$ maps the fraction F of agents choosing option 1 to the probability $r_i(F)$ of that agent choosing 1.

For simplicity of development we focus on the deterministic case, in which agent i chooses option 0 if $F < q$ and option 1 if $F \geq q$, where q is i's threshold. The stochastic case, in which the *most likely* option switches as the agent threshold is crossed, can be handled using a similar derivation. Consider the way the fraction of individuals choosing option 1, F(t), evolves over time. Denote by $r_q(F)$ the agent decision function for agents with threshold q and let g(q) be the probability density function for this threshold. If the agents update their choices *synchronously*, it is easy to show that F(t) evolves according to the discrete-time dynamics

$$F^+ = \int g(q)\, r_q(F)\, dq \equiv r(F),$$

where $r: [0,1] \to [0,1]$ is the average agent decision function for the social context. In the case of positive externality dynamics, r(F) is simply the cumulative density function G(.)

associated with g(.).

Alternatively, if agents update their decisions *asynchronously*, F(t) evolves according to the continuous-time dynamics

$$dF/dt = \lambda(r(F) - F),$$

where $\lambda$ is the rate of agent decision updating. For positive externality dynamics we again have that r(F) is the cumulative density function for the agent decision thresholds.

This basic foundation provides a good starting point for developing useful meso-scale models for positive externality collective dynamics. For example, it is reasonable to suppose that individuals possess varying levels of "inertia" with respect to adopting a new innovation, so that only a fraction $\alpha$ of the agents whose adoption threshold has been exceeded will actually switch from option 0 to option 1 at a given time step. The discrete-time and continuous-time models derived above are easily extended to incorporate this property of social dynamics:

$$F^+ = F + \alpha(G(F) - F),$$
$$dF/dt = \lambda^*(G(F) - F),$$

where G(.) is the cumulative density function for the distribution of the threshold q across the population, as before, and $\lambda^*$ is an $\alpha$-dependent rate constant. Models of this form are derived using a different approach in [3], where they are shown to provide good agreement with empirical data for the diffusion of innovations within social contexts.

The basic model can be extended in other ways. In many applications it is important to capture 1.) the notion that decisions to adopt or not may depend upon *encountering* individuals who have or haven't adopted and 2.) the fact that the likelihood of encountering a particular type, say an adopter, depends on the fraction of this type in the population. For instance, these considerations are found to be important in social movements, in which individuals are faced with the decision of whether to join the movement and, if they join, whether to remain members [4]. Additionally, it should be noted that the basic model describes the *expected* evolution of the social process and is thus deterministic. It is frequently of interest to derive stochastic representations for the social dynamics; this is of particular interest in predictive analysis, where a probabilistic description of process uncertainty can be valuable. Again, these extensions are readily incorporated within the basic model (here we provide the extension for the continuous time case):

$$\Sigma_H: \quad dP/dt = -\beta PM - (\beta PM)^{1/2}\eta_1(t),$$
$$dM/dt = \beta PM + (\beta PM)^{1/2}\eta_1(t) - \delta_1 M - (\delta_1 M)^{1/2}\eta_2(t) - \delta_2 ME - (\delta_2 ME)^{1/2}\eta_3(t),$$
$$dE/dt = \delta_1 M + (\delta_1 M)^{1/2}\eta_2(t) + \delta_2 ME + (\delta_2 ME)^{1/2}\eta_3(t),$$

where P, M, and E denote the fractions of potential members, members, and ex-members of the adopting group, $\beta$, $\delta_1$, and $\delta_2$ are nonnegative constants, and the $\eta_i(t)$ are appropriate random processes [e.g., 13]. A deterministic model of this basic form, derived in a different way, is discussed in [4] and therefore we denote the model $\Sigma_H$. This deterministic version is shown in [4] to provide a useful description of the *local* growth of various social movements. Additional extensions to the basic model are derived in [10].

The basic meso-scale model and its extensions describe the way individual agent decision functions interact to produce collective behavior *within* social contexts. Individuals also interact with those from other contexts, of course, and receive information through channels that transmit to many contexts simultaneously (e.g., mass media). These inter-context interactions and "global" social signals are quantified at the macro-scale level of the multi-scale modeling framework. The basic idea is natural and straightforward. As indicated in Figure 1, we model interdependence between contexts with a graph $G_{sc} = \{V_{sc}, E_{sc}\}$, where each vertex $v \in V_{sc}$ is a social context and each directed edge $e = (v,v') \in E_{sc}$ represents a potential inter-context interaction. More specifically, an edge $(v,v')$ indicates that an agent in context $v'$ can receive decision-relevant information from one in context v. The way agents act upon this information is specified by their decision functions $r_i$. The broadcast of global social signals to individuals is modeled as a context-dependent input $u_v$ to each individual in context v. Thus $G_{sc}$ and the $u_v$ define the macro-scale model structure.

A key task in deriving a macro-scale model is specifying the topology of $G_{sc}$, as this graph encodes the social network structure for the given social process. We employ two approaches to constructing this graph: demographics-based and social network-based. In the former, demographics data are used to define both the contexts themselves (e.g., families, physical neighborhoods) and their proximity. The basic idea is familiar: individuals belong to social groups, which in turn belong to "groups of groups", and so on, giving rise to a hierarchical organization of contexts. For instance, in academics, research groups often belong to academic departments, which are organized into colleges, which in turn form universities, and so on. The proximity of two contexts is specified by their relationship within the hierarchy, and this distance defines the likelihood that individuals from the two contexts will interact. Thus two members of different research units, for example, are more likely to interact if the units belong to the same department than if they are merely part of the same university. The probability of inter-context interaction can, in turn, be used to define the social context graph $G_{sc}$. For example, an edge $(v,v') \in E_{sc}$ can be defined to reflect the interaction probability for individuals from contexts v and v' (in the simplest case, an edge can indicate that this probability exceeds a certain threshold). Alternatively, other forms of "demographics-like" data can be used to infer context relationships. This possibility is illustrated below in an example involving the propagation of the SARS virus.

Another approach to building the social context graph $G_{sc}$ is to infer contexts directly from social network data. One useful method is to define social contexts to correspond to

*graph communities*, that is, sets of individuals in a social network with intra-group edge densities that are significantly higher than expected [12]. Interpreting social network communities in this way has a sound social science basis (see, e.g., [11,12] and the references therein), and numerous fast algorithms exist for detecting graph communities in large-scale networks [12]. Therefore the main challenge with this method for building social context graphs is obtaining social network data for the system of interest. While this task is certainly nontrivial, such data has become much more available in the past decade. For instance, social relationships and interactions increasingly leave "fingerprints" in electronic databases, making convenient the acquisition, manipulation, storage, and analysis of these records.

### C. S-HDS model formulation

We now show that the stochastic hybrid system formalism provides a rigorous, tractable, and expressive framework within which to represent multi-scale social dynamics models. Consider the following

**Definition 2.1:** A *stochastic hybrid dynamical system* (S-HDS) is a feedback interconnection of a continuous-time, continuous state-dependent Markov chain $\{Q, \Lambda(h(x))\}$ and a collection of stochastic differential equations indexed by the Markov chain state q:

$$\Sigma_{\text{S-HDS}} \quad \begin{array}{c} \{Q, \Lambda(h(x))\}, \\ dx = f_q(x,p)dt + G_q(x,p)dw, \\ k = h(x), \end{array}$$

where $q \in Q$ is the discrete state, $x \in X \subseteq \Re^n$ is the state of the continuous system, $p \in \Re^p$ is the vector of system parameters, $\Lambda(x)$ is the matrix of (x-dependent) Markov chain transition rates, $\{f_q\}$ and $\{G_q\}$ are sets of vector and matrix fields characterizing the continuous system dynamics, w is an m-valued Weiner "disturbance" process, and h defines a partition of the continuous state space into subsets labeled with index k.

We now develop an S-HDS representation for an important instantiation of the proposed multi-scale model for social dynamics. It is assumed that:

- the social system consists of N agents distributed over M social contexts;
- agents make binary decisions, choosing from a set O = {0, 1} of options;
- intra-context interactions are fully mixed;
- inter-context interactions involve migration of agents from one context to another;
- the model includes both probabilistic and set-bounded uncertainty descriptions.

The phenomenon of interest is the *diffusion of innovations*, in which an innovation of some kind (e.g., a new technology or idea) is introduced into a social system, and individuals may learn about the innovation from others and decide to adopt it [3]. By definition an innovation is "new", and therefore it is supposed that initially only one or a few of the social contexts have been exposed to it. It is of interest to characterize the extent to which the innovation will spread through the social system.

We propose to model diffusion of innovations as follows:

**Definition 2.2:** The *multi-scale diffusion of innovations S-HDS model* is a tuple

$$\Sigma_{\text{S-HDS, diff}} = \{G_{sc}, Q \times X, \{f_q(x), G_q(x), H_q(x)\}_{q \in Q}, \text{Par}, W, U, \{Q, \Lambda(x)\}\},$$

where

- $G_{sc} = \{V_{sc}, E_{sc}\}$ is the social context graph;
- $Q \times X$ is the system state set, with Q and $X \subseteq \Re^n$ denoting the (finite) discrete and (bounded) continuous state sets, respectively;
- $\{f_q(x), G_q(x), H_q(x)\}_{q \in Q}$, Par, W, U is the S-HDS continuous system, a family of stochastic differential equations which characterizes the intra-context dynamics via vector field/matrix families $\{f_q\}, \{G_q\}, \{H_q\}$, system parameter vector $p \in \text{Par} \subseteq \Re^p$, and system inputs $w \in W \subseteq \Re^m$, $u \in U \subseteq \Re^r$;
- $\{Q, \Lambda(x)\}$ is the S-HDS discrete system, a continuous-time Markov chain which defines inter-context interactions via state set Q and transition rate matrix $\Lambda(x)$.

The social context graph $G_{sc}$ defines the feasible context-context innovation diffusion pathways: if $(v,v') \notin E_{sc}$ then it is not possible for the innovation to spread directly from context v to context v'. The discrete state set $Q = \{0,1\}^M$ specifies which contexts contain at least one adopter of the innovation by labeling such contexts with a '1' (and a '0' otherwise). Thus, for example, state $q = [1\ 0\ 0\ \ldots\ ]^T$ indicates that context 1 has at least one adopter, contexts 2 and 3 have no adopters, and so on. The continuous state space X has coordinates $x_{ij} \in [0,1]$, where $x_{ij}$ is the ith state variable for the continuous system dynamics evolving in context j. For consistency we use the first coordinate for each context, $x_{1j}$, to refer to the fraction of (option 1) adopters for that context. The continuous system dynamics is defined by a family of q-indexed stochastic differential equations $\{\Sigma_{\text{cs, q}}\}_{q \in Q}$, with

$$\Sigma_{\text{cs, q}}: \quad dx = f_q(x,p)dt + G_q(x,p)dw + H_q(x,p)du,$$

where $w \in W$ is a standard Weiner process and $u \in U$ is the exogenous input. Ordinarily w is interpreted as stochastic "noise", while u is employed to represent influences from "global" sources such as mass media. These dynamics quantify intra-context diffusion of the innovation of interest, for instance through models of the form $\Sigma_H$. The Markov chain matrix $\Lambda(x)$ specifies the transition rates for discrete state transitions $q \to q'$ and depends on both $G_{sc}$ and x (e.g., the likelihood that context v will "infect" other contexts depends upon the fraction of adopters in v).

The implementation and performance of the diffusion of innovations model $\Sigma_{\text{S-HDS, diff}}$ are illustrated with several example problems in [10]. The next section highlights the basic aspects of the model by considering a familiar application: large-scale epidemic modeling.

### III. CASE STUDY: SARS EPIDEMIC

The outbreak and rapid spread of severe acute respiratory syndrome (SARS) in 2002-2003 provides a good introductory example of how the proposed S-HDS multi-scale modeling framework can be used to obtain insights regarding important social phenomena. Although the focus of our modeling framework is social dynamics, the framework is easily modified to enable modeling and simulation of the transmission of an infectious disease through society. We now briefly summarize the development and implementation of an S-HDS model for the global SARS epidemic. Additional details concerning the basic model employed in this case study are given in [10]. We begin with the model structure provided by our S-HDS model for innovation diffusion, $\Sigma_{\text{S-HDS, diff}} = \{G_{sc}, Q \times X, \{f_q(x), G_q(x), H_q(x)\}_{q \in Q}, \text{Par}, W, U, \{Q, \Lambda(x)\}\}$. Consider the stochastic susceptible-infected-removed (SIR) model for disease transmission in a fully mixed population [e.g., 14]:

$$\Sigma_{SIR} \quad \begin{aligned} dS/dt &= -\beta SI - (\beta SI)^{1/2} \eta_1(t), \\ dI/dt &= \beta SI + (\beta SI)^{1/2} \eta_1(t) - \delta I - (\delta I)^{1/2} \eta_2(t), \\ dR/dt &= \delta I + (\delta I)^{1/2} \eta_2(t), \end{aligned}$$

where S, I, and R denote the concentrations of agents in the susceptible, infected, and removed states, respectively, $\beta$ and $\delta$ are nonnegative constants, and the $\eta_i(t)$ are appropriate random processes. We implement the SIR model $\Sigma_{SIR}$ on each social context in the multi-scale model; thus $\Sigma_{SIR}$ defines the components $X, \{f_q(x), G_q(x), H_q(x)\}_{q \in Q}, \text{Par}, W, U$ which make up the continuous system portion of the model $\Sigma_{\text{S-HDS, diff}}$.

To obtain a very simple representation for global transmission of the SARS epidemic, we model each country as a social context and assume inter-context interactions consist of individuals traveling between countries along commercial air travel routes (data on air travel was obtained from [15]). This characterization of inter-context dynamics enables the social context graph $G_{sc}$ and the S-HDS discrete system $\{Q, \Lambda(x)\}$ to be constructed directly from publicly available data on airline route topology and passenger traffic density.

The proposed model is very simple, easy to construct, and efficient to simulate (e.g., it runs in minutes on a laptop). Nevertheless, the model provides a useful description for the spread of SARS. For example, Figure 2 shows that simulations of the model are in good agreement with the actual spread of the SARS epidemic during 2002-2003. This example suggests that inter-context interactions can be a key element of disease propagation and that a simple S-HDS model offers an effective way to capture these dynamics.

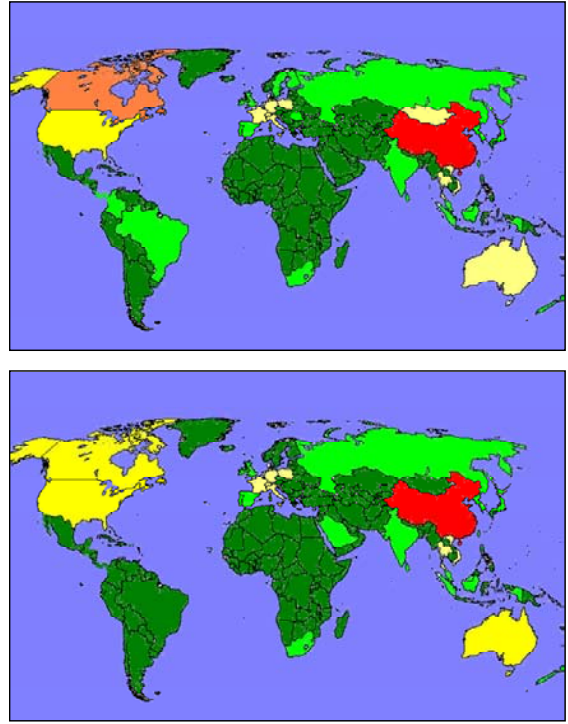

Fig. 2. Geographic spread of SARS epidemic: actual (top) and simulated (bottom) cumulative SARS virus infection levels by country (dark green is lowest, red is highest).